\documentclass[prl,aps,twocolumn,%
showpacs,preprintnumbers,amsmath,amssymb,nofootinbib]{revtex4-2}
\usepackage[utf8]{inputenc}

\usepackage{graphicx}
\usepackage{bm}
\usepackage[breaklinks, colorlinks=true, pdfstartview=FitV, citecolor=blue, urlcolor=blue]{hyperref}
\usepackage{slashed}
\usepackage{array}
\usepackage{mathtools}

\graphicspath{{./Figures/}}

\newcommand{\diff}{\mathrm{d}}

\newcommand{\Diff}{{\mathcal{D}}}

\newcommand{\im}{\mathrm{i}}

\newcommand{\calA}{\mathcal{A}}

\newcommand{\calH}{\mathcal{H}}

\newcommand{\calV}{\mathcal{V}}

\newcommand{\rme}{\mathrm{e}}

\newcommand{\Z}{\mathbb{Z}}
\newcommand{\R}{\mathbb{R}}

\newcommand{\Q}{\mathbb{Q}}
\newcommand{\CP}{\mathbb{C}P^1}
\renewcommand{\hom}{\mathrm{Hom}}

\begin{document}

\preprint{YITP-22-126}

\title{Solitonic symmetry beyond homotopy: Invertibility from bordism and noninvertibility from topological quantum field theory}

\author{Shi Chen}
\email{s.chern@nt.phys.s.u-tokyo.ac.jp} 
\affiliation{Department of Physics, The University of Tokyo,
7-3-1 Hongo, Bunkyo-ku, Tokyo 113-0033, Japan}

\author{Yuya Tanizaki}
\email{yuya.tanizaki@yukawa.kyoto-u.ac.jp}
\affiliation{Yukawa Institute for Theoretical Physics,  Kyoto University, Kyoto 606-8502, Japan}

\begin{abstract}
Solitonic symmetry has been believed to follow the homotopy-group classification of topological solitons.
Here, we point out a more sophisticated algebraic structure when solitons of different dimensions coexist in the spectrum.
We uncover this phenomenon in a concrete quantum field theory, the $4$d $\mathbb{C}P^1$ model.
This model has two kinds of solitonic excitations, vortices and hopfions, which would follow two $U(1)$ solitonic symmetries according to homotopy groups.
Nevertheless, we demonstrate the nonexistence of the hopfion $U(1)$ symmetry by evaluating the hopfion charge of vortex operators.
We clarify that what conserves hopfion numbers is a non-invertible symmetry generated by 3d spin topological quantum field theories (TQFTs).
Its invertible part is just $\mathbb{Z}_2$, which we recognize as a spin bordism invariant.
Compared with the 3d $\mathbb{C}P^1$ model, our work suggests a unified description of solitonic symmetries and couplings to topological phases.
\end{abstract}

\maketitle
\paragraph{Introduction}

Solitons are nonperturbative excitations in quantum field theories (QFTs) and they appear in various areas of contemporary physics, especially almost ubiquitous in high-energy and condensed-matter physics.
Their stability is protected by topological conservation laws and summarised as solitonic symmetries.
It is widely accepted that these conservation laws are classified by the homotopy groups of the target space~\cite{Mermin:1979zz, Coleman:1985rnk, Manton:2004tk}. 
In this Letter, we revisit the solitonic conservation laws in QFTs that allow more than one type of solitonic excitations.
We shall reveal that higher-dimensional solitons may contaminate the conservation law of lower-dimensional solitons, destroy the simple solitonic symmetry predicted by homotopy groups, and lead us to a far more sophisticated symmetry structure.

This phenomenon may be best exemplified by the nonlinear sigma model with the target space $\CP$, i.e. $S^2$, in the $4$d spacetime.
This model is also quite interesting on its own since it arises as an infrared effective description of many condensed-matter and high-energy systems resulting from the spontaneous breaking of an $SU(2)$ symmetry into $U(1)$.
There are two types of solitonic excitations according to the following homotopy groups,
\begin{equation}
    \pi_{1}\left(\CP\right) \simeq 0\,,\quad \pi_{2}\left(\CP\right) \simeq \Z\,,\quad \pi_{3}\left(\CP\right) \simeq \Z\,.
\end{equation}
On the one hand, $\pi_3(\CP)$ classifies the solitons of dimension $1$, i.e. particle excitations, which are often called hopfions (or Hopf solitons)~\cite{Faddeev:1996zj, Gladikowski:1996mb}.
The integer-valued hopfion charge implies a $U(1)$ hopfion symmetry as the topological conservation law.
On the other hand, $\pi_2(\CP)$ classifies the solitons of dimension 2, i.e. stringy excitations, which we shall call vortices.
Similarly, the integer-valued vortex charge implies a $U(1)$ vortex symmetry, except that it is a 1-form symmetry~\cite{Gaiotto:2014kfa}.

As known for a long, unlike the vortex $U(1)$ symmetry, the hopfion $U(1)$ symmetry does not have a local conserved current  although it is a continuous symmetry, which always confused physicists.
As we shall see, this hopfion $U(1)$ is an illusion caused by the homotopy group and does not exist at all.
The authentic hopfion symmetry is merely $\Z_2$, provided that we insist on the reversibility of a symmetry transformation.
This invertible hopfion symmetry follows a bordism classification instead of a homotopy-group classification.
The bordism classification appears repetitively in contemporary physics and notable examples include the classification of invertible topological phases and thereby 't Hooft anomalies~\cite{Wen:2013oza, Kapustin:2014tfa, Freed:2016rqq, Yonekura:2018ufj, Garcia-Etxebarria:2018ajm, Witten:2019bou}.

However, this $\Z_2$ hopfion symmetry provides only an inadequate conservation law.
To obtain a complete conservation law, we must give up the conventional doctrine on the invertibility of symmetry transformations.
Such non-invertible symmetries were recently noticed in 2d QFTs (see \cite{Verlinde:1988sn, Frohlich:2006ch, Bhardwaj:2017xup, Chang:2018iay, Thorngren:2019iar, Komargodski:2020mxz, Nguyen:2021naa}) and have also been recognized in higher dimensions since last year (see e.g.~\cite{Nguyen:2021yld, Koide:2021zxj, Choi:2021kmx, Kaidi:2021xfk, Roumpedakis:2022aik, Hayashi:2022fkw, Bhardwaj:2022yxj, Kaidi:2022uux, Choi:2022jqy, Cordova:2022ieu, Choi:2022rfe}).
They are found to provide additional selection rules and to impose novel constraints on the dynamics.
As we shall see, the complete hopfion symmetry is indeed a non-invertible symmetry.

A remarkable feature of the complete hopfion symmetry is that its symmetry operators are given by 3d topological QFTs (TQFTs).
Like that bordisms can describe invertible topological phases, i.e. short-range entangled topological phases, TQFTs can describe topological orders, i.e. long-range entangled topological phases.
If we reduce the $\CP$ model to 3d spacetime, hopfions become 0-dimensional objects, i.e. instantons.
The $\Z_2$ invertible hopfion symmetry reduces to the discrete $\theta$-angles and the full non-invertible hopfion symmetry reduces to the couplings of the $3$d $\mathbb{C}P^1$ sigma model to topological orders~\cite{Freed:2017rlk, Kobayashi:2021qfj}.
Therefore, our work suggests that solitonic symmetry in a higher-dimensional spacetime and couplings with topological phases in a lower-dimensional spacetime are described by a unified language.
Coupling a topological phase might be phrased as coupling a ``0-form background gauge field'' to the ``(-1)-form solitonic symmetry''.
Such unification may pave the way for further insights into both the solitonic symmetry and the topological phase, as well as their interplay.

\vspace{0.5em}
\paragraph{Operators and solitonic charges}

Hopfions are particles and thus created/annihilated by local operators.
We denote by $B_m(x)$ an operator that creates $m$ hopfions at the location $x$ in the spacetime.
This hopfion operator $B_m(x)$ is characterized as the point defect such that the field configuration on the infinitesimal $S^3$ that surrounds $x$ belongs to the deformation class $m\in\pi_3(\CP)\simeq\Z$.
Similarly, vortices are created/annihilated by line operators because they have one more dimension than hopfions.
We denote by $A_n(M_1)$ an operator that creates $n$ vortices bounded by the loop $M_1$.
This vortex operator $A_n(M_1)$ is characterized as the line defect such that the field configuration on an infinitesimal $S^2$ that links $M_1$ belongs to $n\in\pi_2(\CP)\simeq\Z$.

We are going to look carefully into the solitonic charges of these solitonic operators $B_m$ and $A_n$.
For our purpose, it is convenient to describe the $\CP$ target space using the unit $\mathbb{C}^2$ vector, $\vec{z}=(z_1,z_2)$, with the $U(1)$ gauge redundancy, $\vec{z}(x)\sim \rme^{\im \alpha(x)}\vec{z}(x)$.
We then introduce an auxiliary $U(1)$ gauge field,
\begin{equation}
    a\equiv\im \vec{z}^{\,\dagger} \cdot \diff \vec{z}.
    \label{eq:U1_gauge_field}
\end{equation}
Its properly normalized $U(1)$ curvature $\diff a/2\pi$ is conserved due to the Bianchi identity and integrates to an integer on any closed 2-manifold $M_2$,
\begin{equation}\label{eq:V-charge}
    \int_{M_2}\frac{\diff a}{2\pi}\in \Z\,.
\end{equation}
This integral clearly measures the vortex charge $\pi_2(\CP)$ and $\diff a/2\pi$ gives the 2-form conserved current of the vortex $U(1)$ $1$-form symmetry.
We may construct a vortex $U(1)$ $1$-form symmetry operator $\calV_{\beta}(M_2)$ on $M_2$ through
\begin{equation}\label{eq:gene_vort}
    \calV_{\beta}(M_2)\equiv \exp\left(\im\beta\int_{M_2}\frac{\diff a}{2\pi}\right),\qquad \beta\in\frac{\R}{2\pi\Z}\,.
\end{equation}
If we link $\calV_{\beta}$ with the vortex operator $A_{n}$, we will obtain the phase $\exp(\im\beta n)$.

How to measure the hopfion charge is less obvious.
$\pi_3(\CP)$ comes from $\pi_3(S^3)$ after gauging out a $U(1)$ action and the conserved current for $\pi_3(S^3)$ is easy to find.
Via this trick, the hopfion charge is expressed as the integral of a properly normalized Chern-Simons form~\cite{Whitehead, Wilczek:1983cy, Wu:1984kd},
\begin{equation}\label{eq:Hopf_S3}
    \int_{S^3}\frac{a\diff a}{4\pi^2}\in \Z,
\end{equation}
which is called the Hopf invariant\footnote{For generic $U(1)$ gauge fields, the Chern-Simons form can integrate to any real numbers.
    But our $a$ is restricted to the form of Eq.~\eqref{eq:U1_gauge_field}.
}. 
We may construct a hopfion $U(1)$ symmetry operator $\calH_{\alpha}(S^3)$ on $S^3$ through
\begin{equation}\label{eq:Gene_S3}
    \calH_{\alpha}(S^3)\equiv \exp\left(\im\alpha\int_{S^3}\frac{a\diff a}{4\pi^2}\right),\qquad \alpha\in\frac{\R}{2\pi\mathbb{Z}}\,.
\end{equation}
If we link $\calH_{\alpha}(S^3)$ with the hopfion operator $B_{m}$, we will obtain the phase $\exp(\im\alpha m)$.
From this observation, one might be tempted to define the 3-form current for the hopfion $U(1)$ symmetry as the Chern-Simons form $a\diff a/4\pi^2$.
It is indeed a conserved current since Eq.~\eqref{eq:U1_gauge_field} orders $(\diff a)^2=0$.
However, it is not gauge invariant and thus not a physically sensible current.
As a result, its integral on a general closed $3$-manifold $M_3$ may suffer from gauge ambiguity.

One particularly interesting case is the hopfion charge of the vortex operator $A_n(M_1)$. 
As $M_1$ is a loop, we can surround it by $M_3\simeq S^2\times S^1$.
Let us parameterize the loop $S^1$ by $\tau\in\mathbb{R}/2\pi\Z$ and perform the large gauge transformation, $z\to z'=\rme^{-\im k\tau}z$ with an integer $k$.
Then the auxiliary $U(1)$ gauge field $a\to a'=k\diff \tau+a$.
After this transformation, the integral of the Chern-Simons form changes according to
\begin{align}
    \int_{S^2\times S^1}\frac{a' \diff a'}{4\pi^2} - \int_{S^2\times S^1}\frac{a\diff a}{4\pi^2} = 2nk\,.
\end{align}
Therefore, the hopfion charge of $A_{n\neq0}(M_1)$ are only well-defined modulo $2|n|$,
\begin{equation}\label{eq:Hopf_S2xS1}
    \int_{S^2\times S^1}\frac{a\diff a}{4\pi^2} \in \Z_{2|n|}\,,
\end{equation}
and a gauge-invariant symmetry operator $\calH_{\alpha}(M_3)$ surrounding $A_{n\neq0}(M_1)$ has more restrictive coefficients:
\begin{equation}\label{eq:Gene_S2xS1}
    \calH_{\frac{q}{n}\pi}(S^2\!\times\!S^1)\!\equiv\exp\left(\im\frac{q}{n}\!\int_{S^2\!\times\!S^1}\!\frac{a\diff a}{4\pi}\right)\!,\quad q\in\Z_{2|n|}.
\end{equation}
It means a $\Z_{2|n|}$ symmetry instead of $U(1)$.

The existence of a mod-$2|n|$ Hopf invariant suggests that the vortex charge does not completely constrain a vortex operator:
$A_{n\neq0}$ has $2|n|$ finer deformation classes\footnote{
This fact can be proved via traditional techniques from algebraic topology; see e.g.~\cite{pontrjagin:1941}.
}
which can be distinguished by the mod-$2|n|$ Hopf invariant.
From now on, we shall write a vortex operator as $A_{n,\ell}$ adding a new subscript $\ell\in\Z_{2|n|}$ to label the finer deformation classes.
As an example, we now give an explicit construction of $A_{1,\ell}$ with $\ell=0,1$.
In general, a line defect on $M_1$ is characterized by the boundary condition of the $\CP$ field on the infinitesimal $M_3\simeq S^2\!\times \!S^1$ that surrounds $M_1$.
We can view it as a family of $\CP$ configurations on $S^2$ parametrized by $S^1$.
For $A_{1,\ell}(M_1)$, this $S^1$-parametrization simply describes the rotation process of a $2$-sphere.
Recall that $\pi_1(SO(3))\simeq \Z_2$, which implies there are two deformation classes of such rotation processes.
The untwisted rotation process corresponds to $A_{1,0}$ and the twisted rotation process corresponds to $A_{1,1}$.

\vspace{0.5em}
\paragraph{Invertible solitonic symmetry from bordism}

We have encountered a strange situation.
The range of the hopfion charge depends on what operator it measures [see Eqs.~\eqref{eq:Hopf_S3} and~\eqref{eq:Hopf_S2xS1}] and, accordingly, the structure of the hopfion symmetry depends on what operator it acts on [see Eqs.~\eqref{eq:Gene_S3} and~\eqref{eq:Gene_S2xS1}].
We point out that only the $\mathbb{Z}_2$ symmetry generator, 
\begin{equation}\label{eq:Gene_M3}
    \calH_{\pi}(M_3)\!\equiv \exp\left(\im \int_{M_3}\frac{a\diff a}{4\pi}\right)\ \to\ \pm1\,,
\end{equation}
is always well-defined for all these cases.
It is then natural to guess that the true hopfion symmetry is $\Z_2$ rather than $U(1)$.
To demonstrate this statement, we now describe some correlation functions that violate the $U(1)$ selection rule but are consistent with the $\Z_2$ selection rule.

\begin{figure}[t]
    \centering
    \includegraphics[width=0.7\columnwidth]{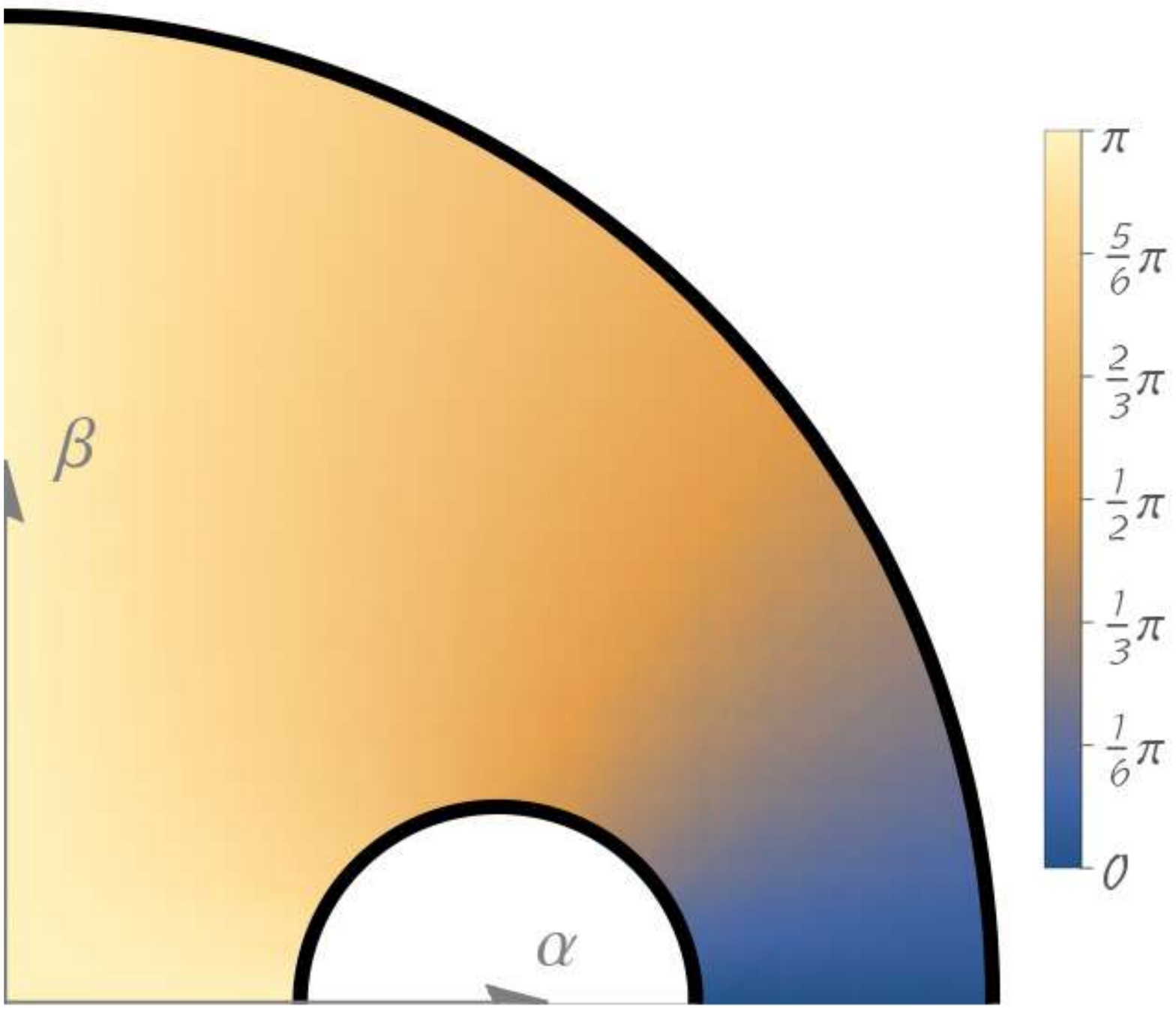}
    \caption{The range of the coordinates $\alpha,\beta$ defined in Eq.~\eqref{eq:M_4} and the function $\theta(\alpha,\beta)$ defined in Eq.~\eqref{eq:theta}.
    The solid quarter arc describes the point defect and the solid semicircle describes the line defect. 
    On the $\alpha$-axis and the $\beta$-axis, $\theta$ is either $0$ or $\pi$, which ensures that the $\mathbb{C}P^1$ configuration is regular and single-valued. }
    \label{fig:R}
\end{figure}

Let us evaluate $\langle A_{1,\ell}(M_1) B_m(x)\rangle$.
We put a line defect on $M_1$ and a point defect at $x$ in the (infrared-regularized) spacetime $S^4$.
The spacetime with the singularities caused by defects removed, denoted by $M_4$, can be described as follows.
First, we consider a system of coordinates $(\alpha,\beta,\mu,\nu)$ on $\mathbb{C}^2$ via
\begin{equation}\label{eq:M_4}
    z_1 = \alpha\rme^{\im\mu}\,,\qquad
    z_2 = \beta\rme^{\im\nu}\,.
\end{equation}
We require $\alpha,\beta\geq0$ and $\mu,\nu\in\mathbb{R}/2\pi\Z$. 
Then, we define a subregion by the following inequalities,
\begin{equation}
    \alpha^2 + \beta^2 \leq (2c)^2\,,\qquad (\alpha-c)^2+\beta^2\geq d^2\,,
\end{equation}
where $c$ and $d$ are constants such that $0<d<c$. 
The $\alpha$-$\beta$ quadrant bounded by these inequalities is shown in Fig.~\ref{fig:R}. 
The subregion in $\mathbb{C}^2$ constrained by these inequalities is exactly diffeomorphic to $M_4$.
We now write down a series of concrete $\CP$ configurations on $M_4$.
For this purpose, we take the standard spherical coordinates $(\theta,\varphi)$ on $\CP$ (recall that $0\le\theta\le\pi$ and $\varphi\in\mathbb{R}/2\pi\Z$).
For each integer $m$, we introduce a configuration
$\phi_m:M_4\mapsto\CP$ defined by $\phi_m(\alpha,\beta,\mu,\nu)\equiv\bigl(\theta(\alpha,\beta),\ \varphi_m(\mu,\nu)\bigr)$ where
\begin{subequations}
\begin{align}
    \theta(\alpha,\beta) &\equiv \mathrm{Arg}[(\alpha+\im\beta)^2 - c^2]\,,\label{eq:theta}\\
    \varphi_m(\mu,\nu) &\equiv \nu + m\mu\,.
\end{align}
\end{subequations}
We plot $\theta(\alpha,\beta)$ in Fig.~\ref{fig:R} for clarity.
Our $\phi_m|_{\alpha^2+\beta^2=(2c)^2}:S^3\mapsto\CP$, i.e. the configuration restricted on the $S^3$ part of $\partial M_4$, describes nothing but the point defect $B_m$.
Different $\phi_m$'s describe different deformation classes of point defects.
But this is not the case for line defects, which are described by $\phi_m|_{(\alpha-c)^2+\beta^2=d^2}:S^2\times S^1\mapsto\CP$.
All the $\phi_m|_{(\alpha-c)^2+\beta^2=d^2}$ with even $m$ give the same deformation class $A_{1,0}$, and all the $\phi_m|_{(\alpha-c)^2+\beta^2=d^2}$ with odd $m$ give $A_{1,1}$, as a consequence of $\pi_1(SO(3))\simeq \mathbb{Z}_2$.
We can recognize this fact via the explicit description of $A_{1,\ell}$ we introduced before.
Therefore, $\phi_m$'s provide us with bordisms connecting $A_{1,m\,\mathrm{mod}\,2}$ and $B_m$, the presence of which proves the following results,
\begin{equation}\label{eq:corre}
\begin{split}
    &\langle A_{1,0}(M_1) B_{m}(x)\rangle \neq 0\quad\text{for even }m\,,\\
    &\langle A_{1,1}(M_1) B_{m}(x)\rangle \neq 0\quad\text{for odd }m\,.
\end{split}
\end{equation}
These correlation functions show that $A_{1,0}$ emits/absorbs any even number of hopfions while $A_{1,1}$ emits/absorbs any odd number of hopfions.
$B_m$ and $B_{m+2}$ have to share the same hopfion charge, provided the invertibility.
This intriguing phenomenon is consistent with the $\Z_2$ selection rule described by Eq.~\eqref{eq:Gene_M3} but violates the $U(1)$ selection rule predicted by homotopy groups.

As we have just seen, the evaluation of the correlation functions of defect operators is actually a bordism problem.
Our $\phi_m$ precisely establishes a bordism between a configuration on $S^3$ and another on $S^2\!\times\!S^1$.
Actually, the $\Z_2$ generator~\eqref{eq:Gene_M3} detects a reduced spin bordism group,
\begin{align}
    \widetilde{\Omega}_3^{\mathrm{spin}}(\CP)\simeq \Z_2\,,
\end{align} 
and thereby generates $\hom_{\Z}(\widetilde{\Omega}_3^{\mathrm{spin}}(\CP), U(1))\simeq\Z_2$.
Thus the symmetry operator can be defined on any closed spin 3-manifold.
This bordism group also appears if we lower one spacetime dimension~\cite{Freed:2017rlk, Kobayashi:2021qfj}.
In that context, our symmetry operator in Eq.~\eqref{eq:Gene_M3} precisely gives the only invertible topological phase that can be coupled to the 3d $\CP$ model~\cite{Freed:2017rlk, Kobayashi:2021qfj}, (i.e. a $\Z_2$ $\theta$-angle).

\vspace{0.5em}
\paragraph{Non-invertible solitonic symmetry from TQFT}

The collection of all operators including $A_{n,\ell}$'s and $B_m$'s follow the $\Z_2$ selection rule.
Nevertheless, if we look at a sub-collection of operators, we can see enhanced selection rules larger than $\Z_2$.
For example, if we consider the collection of $A_{n,\ell}$'s with $n=0\!\!\mod N$, as well as $B_{m}$'s, we shall see a $\Z_{2N}$ selection rule.
More drastically, if we confine ourselves to the collection of mere $B_m$'s, the $U(1)$ seems to revive.
If these selection rules are given by some conserved charges, their algebraic structure cannot be a conventional group-like one.
Such a generalized form of symmetry is called non-invertible symmetry in the literature.

Let us try to construct operators that generates complete hopfion symmetry.
Such an operator has to be a bordism covariant instead of a bordism invariant.
We need an ``intelligent'' functional $\calH_{\alpha}(M_3)$ that yields different types of hopfion symmetry [Eq.~\eqref{eq:Gene_S3} or Eq.~\eqref{eq:Gene_S2xS1}] according to the topology and the vortex charge of $M_3$.
Interestingly, for special values of coefficients, $\alpha\in\pi\Q$, the partition function of a $\nu\!=\!\alpha/\pi$ fractional quantum Hall state exactly provides such an intelligent functional.
Let us start with the simplest case of $\alpha=\frac{\pi}{N}$ for a positive integer $N$.
Then a symmetry operator can be defined as
\begin{equation}\label{eq:Gene_U(1)N}
    \calH_{\frac{\pi}{N}}(M_3) \equiv\!\int\!\Diff b\, \exp\!\left[-\im\!\int_{M_3}\!\left(\frac{N}{4\pi}b\,\diff b + \frac{1}{2\pi}b\,\diff a \right)\right] , 
\end{equation}
where $b$ is a $U(1)$ gauge field defined \textit{only} on $M_3$. 
This auxiliary $U(1)$ gauge field is introduced just to define a TQFT partition function and is integrated out in the path integral~\eqref{eq:Gene_U(1)N}.
This is the partition function of the $U(1)_N$ spin Chern-Simons theory which describes a $\nu\!=\!1/N$ fractional quantum Hall state.

We now demonstrate that Eq.~\eqref{eq:Gene_U(1)N} has the desired feature.
Eq.~\eqref{eq:Gene_U(1)N} is a quadratic integral and thus its phase is given by the classical action of the saddle.
Let $M_3$ be an $S^3$ surrounding a $B_m$.
Then the triviality of $U(1)$ bundles on $S^3$ allows us to recover Eq.~\eqref{eq:Gene_S3}, i.e.,
\begin{align}
    \calH_{\frac{\pi}{N}}(S^3)&\to  \exp\left(\frac{\im (\pi/N)}{4\pi^2}\int_{S^3}a\diff a\right)\notag\\
    &= \exp\left(\im\frac{\pi}{N}m\right)\,,
\end{align}
which detects the integer Hopf invariant, and this observation justifies to call \eqref{eq:Gene_U(1)N} as the hopfion symmetry generator.
Now, let $M_3$ be an $S^2\times S^1$ surrounding an $A_{n,\ell}$.
Then we must very carefully perform the $b$-path-integral due to the nontrivial $U(1)$ bundle caused by the vortex charge.
This path integral vanishes if the vortex charge $n$ is not divisible by $N$ since there is no saddle in this case.
When $N$ divides $n$, Eq.~\eqref{eq:Gene_U(1)N} recovers to Eq.~\eqref{eq:Gene_S2xS1} with $q=n/N$.
This leads us to the following result,
\begin{equation}\label{eq:HopfionActio_vortex}
    \calH_{\frac{\pi}{N}}(S^2\!\times\!S^1)\to
    \begin{cases}
    \exp\left(\im\frac{\pi}{N}\ell\right)\,,\quad n=0\!\!\mod N \\ 
    0\,,\quad n\neq0\!\!\mod N
    \end{cases}.
\end{equation}
This result is precisely invariant under the transformation $\ell\to\ell + 2n$ and thus detects the mod-$2|n|$ Hopf invariant.
Therefore, the symmetry operator in Eq.~\eqref{eq:Gene_U(1)N} is intelligent enough to yield different types of hopfion symmetry according to different situations.

Let us now turn to general rational coefficients.
The $U(1)_N$ Chern-Simons theory we just used belongs to a family called the spin minimal Abelian TQFTs~\cite{Hsin:2018vcg} (they were first studied in~\cite{Moore:1988qv}; see also~\cite{Bonderson:2007ci, Barkeshli:2014cna}).
A member of this family is denoted by $\calA^{N,p}$ for a positive integer $N$ and a mod-$N$ integer $p$ such that $\gcd(N,p)=1$.
We can characterize $\calA^{N,p}$ as the minimal 3d spin TQFT that has a $\Z_N$ 1-form symmetry whose 't Hooft anomaly is labeled by $p$.
Every $\calA^{N,p}$ can be expressed as an Abelian Chern-Simons theory due to its Abelian nature and describes a $\nu\!=\!\frac{p}{N}$ fractional quantum Hall state.
We couple $\diff a/N$ as the $2$-form gauge field of their $\Z_N$ $1$-form symmetry.
Then the hopfion symmetry operators are given by their partition functions, i.e.,\footnote{These operators might not give the complete list of generators for the hopfion symmetry. 
However, they constitute at least a dense subset of the complete list, which thus captures the essential ingredient of the symmetry.}
\begin{subequations}\label{eq:gene_TQFT}
\begin{align}
    \calH_{\frac{p}{N}\pi}(M_3)&\equiv \calA^{N,p}\left[M_3,\diff a/N\right]\,,\\
    \calH_{(\frac{p}{N}+1)\pi}(M_3)&\equiv\calH_{\frac{p}{N}\pi}(M_3)\ \calH_{\pi}(M_3)\,,
\end{align}
\end{subequations}
where $\calH_{\pi}(M_3)$ has been defined in Eq.~\eqref{eq:Gene_M3}.

$\calA^{N,p}$ was recently used to furnish the construction of chiral symmetry in 4d QED~\cite{Choi:2022jqy, Cordova:2022ieu}.
As clarified there, once we accept TQFTs as operators, we indeed obtain non-invertible symmetry and need to abandon the group-like fusion rule.
As a remarkable example, let us consider the fusion of $\calH_{\frac{\pi}{N}}(M_3)$ and its seemingly inverse $\calH_{2\pi-\frac{\pi}{N}}(M_3) = \calH_{\frac{\pi}{N}}^\dagger(M_3)$. Then we obtain the following result instead of the identity:
\begin{equation}\label{eq:CondenOperator}
    \calH_{\frac{\pi}{N}}(M_3)\times \calH_{2\pi-\frac{\pi}{N}}(M_3) = \mathcal{C}(M_3),  
\end{equation}
where $\mathcal{C}(M_3)$ is the condensation operator (see e.g.~\cite{Gaiotto:2019xmp, Roumpedakis:2022aik}) defined via gauging the $\mathbb{Z}_N$ subgroup of the vortex $U(1)$ $1$-form symmetry only on $M_3$. 
We emphasize that the right-hand side of Eq.~\eqref{eq:CondenOperator} cannot be the identity operator, because the left-hand side acts on vortex operators nontrivially as we can readily deduce from Eq.~\eqref{eq:HopfionActio_vortex}.
This clarifies the non-invertible nature of the algebraic structure for the hopfion symmetry. 
The $\Z_2$ symmetry we found before generated by $\calH_{\pi}(M_3)$ is the invertible part of this complete hopfion symmetry.

Like the invertible $\Z_2$ symmetry operators, the non-invertible symmetry operators~\eqref{eq:gene_TQFT} also appear in the 3d spacetime.
In that context, they just describe the couplings of the 3d $\CP$ model to $\calA^{N,p}$.
The particular case of Eq.~\eqref{eq:Gene_U(1)N} was discussed in Ref.~\cite{Freed:2017rlk}.
Due to the minimal nature of $\calA^{N,p}$, the coupling to any topological order must factor through $\calA^{N,p}$.
Therefore, Eq.~\eqref{eq:gene_TQFT} actually classifies all the possible couplings of the 3d $\CP$ model to topological orders.
We thus arrive at a unified description of different phenomena in different spacetime dimensions:
The solitonic symmetry in a higher-dimensional spacetime exactly classifies the couplings with topological phases in a lower-dimensional spacetime.
In particular, the invertible solitonic symmetry corresponds to invertible phases while the non-invertible solitonic symmetry corresponds to topological orders.
A coupling to a topological phase might be phrased as a ``0-form background gauge field'' of the ``(-1)-form solitonic symmetry''.

\vspace{0.5em}
\paragraph{Summary and outlook}

We clarified that hopfions in the 4d $\CP$ model follow an unexpected sophisticated conservation law as a non-invertible symmetry instead of $U(1)$ predicted by $\pi_3(\CP)\simeq\Z$.
The symmetry is generated by spin TQFTs and, in particular, its invertible part $\Z_2$ is generated by spin bordism invariant.
This spin nature of the hopfion symmetry signifies hopfions' capability of being a fermion\footnote{
They are not necessarily fermions.
Statistics of solitons are determined by the topological term in the Lagrangian.
In the invertible case 
$\hom_{\Z}(\widetilde{\Omega}_4^{\mathrm{spin}}(\CP), U(1))\simeq\Z_2$, a trivial $\theta$-angle renders hopfions bosons while a nontrivial $\theta$-angle makes hopfions fermions.
}.

We would like to convey two messages via this work.
First, the homotopy groups might not correctly capture the solitonic symmetry when the spectrum includes solitonic excitations of different dimensions.
Second, two prominent topological phenomena, the solitonic symmetry and the topological phase, are perhaps supposed to be treated in a unified scheme.
We hope these messages will lead us to further insights into the topological phenomena in contemporary physics.
The complete algebraic structure of the solitonic charge/symmetry, something like a ``generalized (co)homology with TQFT coefficients'', awaits to be explored by physicists/mathematicians.

\vspace{0.5em}
\paragraph{Acknowledgements}
This work was initiated during S.~C.'s visit to YITP with the Atom-type visiting program, and the authors appreciate the hospitality of Yukawa institute. 
This work was supported by JSPS KAKENHI Grant No. 21J20877 (S.~C.), 22H01218 and
20K22350 (Y.~T.), and also by the Center for Gravitational Physics and Quantum Information (CGPQI). 

\bibliographystyle{utphys}
\bibliography{./letter}
\end{document}